\RequirePackage{fix-cm}
\documentclass[smallextended]{svjour3}       
\smartqed  
\usepackage{graphicx}
\usepackage{amssymb}
\usepackage{amsmath}
\usepackage{color}
\usepackage{soul}

\newcommand{\red}[1]{{\color{black} #1}}

\begin{document}

\title{Cooper-Pair Tunneling in Small Josephson Junction Arrays under Radio-Frequency Irradiation}

\author{Godwill Mbiti Kanyolo \and Kouichi Takeda \and Yoshinao Mizugaki \and Takeo Kato \and Hiroshi Shimada
}


\institute{G. M. Kanyolo$^{\dagger}$, K. Takeda, Y. Mizugaki, and H. Shimada$^{*}$ \at
Department of Engineering Science, The University of Electro-Communications,\\
1-5-1 Chofugaoka, Chofu, Tokyo 182-8585, Japan \\
\email{gmkanyolo@gmail.com$^{\dagger}$}\\
\email{hiroshi.shimada@uec.ac.jp$^{*}$}           
           \and
           T. Kato \at
              Institute for Solid State Physics, The University of Tokyo,\\ 5-1-5 Kashiwanoha, Kashiwa 277-8581, Japan
}

\date{Received: date / Accepted: date}

\maketitle

\begin{abstract}
The influence of radio frequency microwaves on the Coulomb blockade characteristics in small Josephson junctions was studied using a one-dimensional array of ten small Al tunnel junctions in the frequency range from 1 MHz to 1000 MHz. Coulomb blockade voltage ($V_{\rm th}$) is diminished with increasing microwave power ($V_{\rm ac}$), where the $V_{\rm th}$--$V_{\rm ac}$ plots for varied frequencies fall on a single curve. We observed and theoretically analyzed a magnetic field {\it dependent} renormalization of the applied microwave power, in addition to a magnetic-field {\it independent} renormalization effect explained using an effective circuit approach of the array. Due to its high sensitivity to microwave power, the array is well-suited for on-chip detection applications in low temperature environments.

\end{abstract}
\makeatletter
\def\gtrsim{\mathrel{\hbox{\rlap{\hbox{\lower4pt\hbox{$\sim$}}}\hbox{$>$}}}}
\def\lsssim{\mathrel{\hbox{\rlap{\hbox{\lower4pt\hbox{$\sim$}}}\hbox{$<$}}}}
\makeatother

\section{Introduction}

Since the pioneering theoretical work by Likharev and co-workers \cite{Likharev,Likharev2,Averin1991}, small Josephson junctions have been thought of as a dual system to large Josephson junctions -- the roles of current and voltage are interchanged. In the case of large Josephson junctions, their effective interaction with oscillating electromagnetic fields has intensively been studied, demonstrating their unique suitability for microwave-based applications such as the metrological standard for the \textit{Volt} (in terms of voltage ($V$) Shapiro steps) and other microwave-based devices \cite{Barone}.

Thus, the dual system holds enormous promise for complementary applications such as a metrological standard for the \textit{Ampere} in terms of the current ($I$) Shapiro steps \cite{Likharev,Likharev2}. 
However, due to their mesoscopic size, microwave based studies with oscillating electromagnetic fields face daunting experimental and theoretical challenges. 
In particular, small junctions are prone to quantum and thermal fluctuations, thus a well-known (fluctuation-dissipation) theorem applies \cite {Kubo1966,FDT1951}. Due to their fluctuative nature, the characteristics of small junctions generally cannot be analyzed separate from their dissipative environment \cite{Caldeira-Leggett1983,Leggett1984}. Consequently, special considerations and techniques are required to observe such dual characteristics as the Coulomb blockade and Bloch oscillations in a single small junction \cite{Watanabe2001,Watanabe2003}. For a one-dimensional (1D) array of small Josephson junctions, however, Coulomb blockade is easily observed when the junction parameters \red{such as the tunnel resistance $R_{\rm T}$ and the capacitance $C$ of {\it each} junction} are adequately chosen, such that $E_{\rm J}<E_{\rm c}=e^2/2C$ and $R_{\rm T} > R_{\rm Q}=\hbar/e^2$, since the electromagnetic environment of each junction is dominated by the environment of the rest of the array \cite{Haviland1996}. Here, $E_{\rm J}$ is the Josephson coupling energy of a single junction, and $e$ and $\hbar$ respectively denote the elementary charge and the reduced Planck constant. Thus, one-dimensional arrays are more suitable than single junctions to studying the interaction of small Josephson junctions with radio frequency (RF) electromagnetic fields (microwaves).

Pioneering experimental work by Delsing \cite{Delsing1992} intensively studied the weak Josephson coupling limit $E_{\rm J}/E_{\rm c} < 0.1$ and high tunnel resistances $R_{\rm T}>R_{\rm Q}$, where Cooper-pair tunneling is virtually non-existent and the quasi-particle current is dominant. This work observed the lifting of Coulomb blockade of the quasi-particle current by a thermal bath and external microwaves, in addition to small peak structures in the differential conductance at currents $I_{n}=nef$, with $f$ being the frequency of the microwave and $n$ an integer, corresponding to microwave phase-locked narrow-band single electron tunneling oscillations previously predicted by Likhalev \cite{Likharev3}. 
Recently, Billangeon et al. \cite{Billangeon2007} studied the ac Josephson effect and Landau-Zener transitions by detecting, through photon-assisted quasi-particle tunneling in a superconductor-insulator-superconductor (SIS) junction, the microwave emission by a single Cooper-pair transistor (SCPT) in the Cooper-pair transport dominant regime. Photon-assisted tunneling in a SIS junction was further applied to the parametric amplification of microwaves \cite{Jebari2018,Mendes2019}.
In turn, by using photo-resistance measurement, Liou et al. \cite{Liou2008,Liou2014} studied the modulation of the $I$--$V$ characteristics of a 1D array of small dc superconducting quantum interference devices (SQUIDs) with the application of RF electromagnetic fields (microwaves) in the phase-charge crossover regime. Their SQUID-to-SQUID junction had $E_{\rm J}/E_{\rm c}>5$ and $R_{\rm T}<R_{\rm Q}$ in the absence of magnetic flux through the loop.

A recent theoretical consideration by Grabert \cite{Grabert2015} applies the standard $P(E)$ approach \cite{Ingold_Nazarov1992,Falci1991} to derive an expression for dynamical Coulomb blockade in a normal junction indirectly irradiated by RF electromagnetic field via its environment. New features in the $I$--$V$ characteristics include the modulation of the power reaching the junction (zero-th mode) as well as higher order effects of excitation of the environment impedance electromagnetic modes by the microwaves (harmonics). 
The renormalization of the $I$--$V$ characteristics of a tunnel junction due to the quantum fluctuation of an LC-circuit environment was studied both theoretically and experimentally by Parlavvecchio et al. \cite{Parlavecchio2015}. 
However, these works focus on the single junction.

In this work, we performed a quantitative study of the influence of RF electromagnetic fields applied to the small Josephson junction array with $0.1 < E_{\rm J}/E_{\rm c} < 1$ and $R_{\rm T}>R_{\rm Q}$ on its $I$--$V$ characteristics.
In this parameter range, the transport at low bias is dominantly due to Cooper pairs, and the dual characteristics to the large junction, i.e. the Coulomb blockade of Cooper-pair tunneling, is observed without special arrangement of the external environment \cite{Haviland1996}. We observed the Coulomb blockade of Cooper-pair tunneling gradually lifted with increase in applied power of the microwaves irrespective of its frequency $f$ in the low frequency ranges of 1 MHz $\leq f \leq$ 1000 MHz where $hf\le k_{\rm B}T$. Here, $h$, $k_{\rm B}$ and $T$ are Planck constant, Boltzmann constant and temperature, respectively. 

In addition, since the array is superconducting, the environment also depends on external magnetic fields $H$ via the ratio $E_{\rm J}(H)/E_{\rm c}$ which determines the Bloch energy band of the junction. Suppressing $E_{\rm J}$ with $H$ decreases the band gap and increases the threshold voltage for Cooper-pair tunneling in each junction \cite{Likharev2}.
\red{In the absence of RF field, we apply a significant constant magnetic field, $H$ = 500 Oe, which increases the Cooper pair Coulomb blockade (voltage) width to nearly its maximum value by a factor of about 1.4} \cite{Haviland1996,Shimada2012}. On applying the RF field, we find that the curve of the measured threshold voltage for $H= 500 \ {\rm Oe}$ as a function of the amplitude of the applied ac voltage coincides with that for $H= 0$ after rescaling the voltages with the common factor, 1.4.

By comparing the experimental results to our simulation results with the standard expression of photon-assisted tunneling for low frequencies \cite{Tien-Gordon1963,Hamilton1970} consistent with $P(E)$ theory \cite{Falci1991}, we find that the RF-power detected by the array is renormalized by a factor $\Xi$ less than $1$, that is {\it independent} of applied frequency and magnetic field. 
We conclude that this renormalization originates from a voltage division in an effective series circuit, which is composed of a single junction connected to the voltage source and the combined capacitance of a half of the junction array.
On the basis of this theory, we also estimate the renormalization factor using the relation $\Xi = \exp(-\Lambda^{-1})$, where $\Lambda$ is the soliton length of the array \cite{Bakhvalov1989}. 

These results offer the prospect of using the observed gradual decrease in Coulomb blockade voltage with microwave power for detection of low RF electromagnetic field power in varied (on-chip) environments. This flexibility in use is due to its small-signal sensitivity estimated to be \red{more than $10^6$} V/W.

\section{Experimental Method}
\subsection{Sample}
Since the Coulomb blockade width is expected to scale with the number of junctions \cite{Maibaum2011}, the response of Josephson junction arrays to application of RF microwaves is expected to be more pronounced compared to single junctions.
However, since the photon-assisted tunneling along the array is sensitive to the dynamics of the Cooper-pair solitons instead of the Cooper pairs themselves, analysis of experimental results becomes taxing. 
Choosing to fabricate an array with the number of junctions $N_0$ such that $N_0$ is comparable to the charge
soliton length $\Lambda$ allows one to effectively analyze the photon-assisted tunneling through the array by the single
junction equivalent circuit. We, thus, chose $N_0=$10 based on the expected parameters of samples.

We fabricated the one-dimensional array of ten small Al/AlO${}_{\rm x}$/Al Josephson junctions of cross-sectional area 100 $\times$ 200 nm$^{2}$ with the island electrode of length, $l=1\, \mu {\rm m}$, having the parameters $0.1<E_{\rm J}/E_{\rm c}<1$ and $R_{\rm T}>R_{\rm Q}$ by electron-beam lithography and angle evaporation on an oxidized Si substrate. The thickness of the base and counter Al electrodes was 25 and 40 nm respectively. The measured parameters for the sample are tabulated in Table~\ref{tab:a}. 
$R_{\rm T}$ and $E_{\rm c}$ of the junction were obtained from the asymptotic $I$--$V$ curve at high bias by taking its differential resistance and the offset voltage \cite{Averin1991,Delsing1992}, the superconducting gap $\Delta$ of the Al electrodes for $H=$ 0 and 500 Oe were derived from the differential conductance characteristics ${\rm d}I/{\rm d}V$--$V$ 
and the measured $\Delta$--$H$ dependence,
and $E_{\rm J}$ of the junction was calculated by using the Ambegaokar-Baratoff formula \cite{A-B}.

\begin{table}
\label{table:parameter}
\caption{Average parameters for a single junction in the array. 
         The tabulated parameters are: the tunnel resistance $R_\mathrm{T}$, capacitance $C$, charging energy $E_\mathrm{c}$, superconducting gap $\Delta$ of the Al electrode,
         Josephson coupling energy $E_\mathrm{J}$, and $E_\mathrm{J}$-to-$E_\mathrm{c}$ ratio. The last three parameters are for $H$ = 0 Oe and 500 Oe, where $H$ = 500 Oe is the value of magnetic field with the nearly largest Cooper pair Coulomb voltage gap.}\label{tab:a}
    \begin{center}
        \begin{tabular}{ccccccc}\hline
            $R_\mathrm{T}$ / k${\rm \Omega}$  & $C$ / fF & $E_\mathrm{c} / \mu$eV & $\Delta / \mu$eV & $E_\mathrm{J} / \mu$eV & $E_\mathrm{J}/E_\mathrm{c}$ & $H$ / Oe\\
            \hline\hline
            35.1 & 0.72 & 110 &
            $\begin{array}{@{}c@{}}	165\\
                                        133
            \end{array}$ &
            $\begin{array}{@{}c@{}}	30.3\\
                                        24.5
            \end{array}$ &
            $\begin{array}{@{}c@{}}	0.27\\
                                        0.22
            \end{array}$ &
            $\begin{array}{@{}c@{}}	0\\
                                        500
            \end{array}$\\ 
            \hline
        \end{tabular}
     \end{center}
\end{table}

\subsection{Experimental setup}
\begin{figure}[b]
\begin{center}
\includegraphics[width=0.6\columnwidth,clip=true]{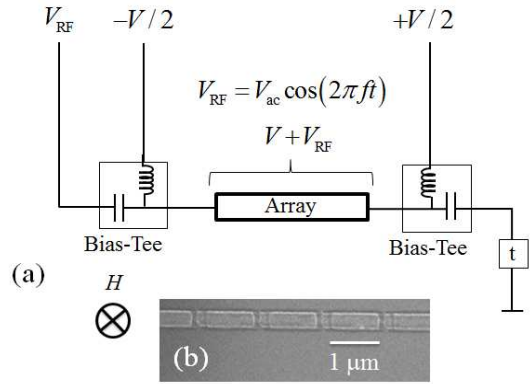}
  \caption{Schematic diagram of the measurement circuit (a) with the scanning electron micrograph (SEM) of the array (b). Both bias-tees combine the RF signal $V_{\rm RF}$ 
           with the symmetrically biasing dc voltage $\pm V/2$ to yield $V + V_{\rm RF}$, with $V_{\rm RF} = V_{\rm ac}\cos (2\pi f t)$ the ac voltage or the applied RF electromagnetic field. \red{The RF line of the bias-tee on the right is terminated with a 50 $\rm \Omega$ terminator "t"}. $H$ denotes the magnetic field, with the circled-cross indicating the applied direction perpendicular to the sample substrate.}
  \label{circuit}
\end{center}
\end{figure}
For the low temperature measurement, we used a typical wet dilution refrigerator with a base temperature of 40 mK.
The refrigerator was fitted with a 50 $\rm \Omega$ semi-rigid coaxial circuitry: a CuNi coaxial cable from the room temperature to the still plate with \red{10 dB and 20 dB} cryogenic attenuators thermally anchored at the 1K-pot and the still plates, respectively, followed by a stainless-steel coaxial cable from the still plate to the mixing chamber (MC) plate where it is connected to a copper coaxial cable via a \red{20 dB} cryogenic attenuator thermally anchored there. The copper coaxial cable is connected to a commercial bias tee, and its output is lead to the copper sample chamber with a copper coaxial cable. Both the bias tee and the sample chamber were anchored at the MC plate. Another bias tee at the MC plate was also connected to the sample chamber in the same way with its RF port terminated with a terminator. The circuitry at the MC plate is schematically depicted in Fig. \ref{circuit}.

As in Fig. \ref{circuit}, the dc and ac signals were combined by a bias-tee. Each dc line goes through low-pass filters with the cut-off frequency of 3.4 kHz at the MC plate to the bias-tee.
The sample was mounted in the copper chamber at the MC plate which had MMCX connectors for the RF circuitry, and the central conductors of the MMCX connectors were directly connected to the Au pads on the sample chip with 4 mm-long Au wires. The on-chip Au lead from the pad to the array was approximately 3 mm long.
The measurements were done in an electromagnetically shielded environment. 
The ac signal, with a frequency range of 1 MHz $\leq f \leq$ 1000 MHz, was generated using a standard network analyzer and applied to the sample.

The dc $I$--$V$ characteristics were measured by the so-called r-bias method \cite{Delsing_thesis}: 
a well-defined resistance $r$ (1 M${\rm \Omega}$ or 5 M${\rm \Omega}$) was inserted in series with the sample, and both were generally voltage-biased; the voltages on $r$ 
and the sample were measured with differential amplifiers with high input impedance to obtain the $I$ and $V$ values; 
the bias voltage was symmetrically applied to the sample with respect to the ground to reduce the noise (Fig. \ref{circuit}). 
To decrease the superconducting gap of the Al electrodes and increase the Coulomb blockade voltage of the array, a magnetic field $H$ was applied perpendicular to the substrate using a superconducting magnet \cite{Haviland1996}.

\subsection{Calibration of the ac input}
The frequency characteristics of the transmission coefficient of the RF line, $\alpha(f)$, was calibrated in advance. To obtain $\alpha(f)$ we prepared two identical transmission lines in the cryostat, each extending from the room temperature terminal to the sample chamber, including attenuators and a bias-tee. We simply shorted the ends of the two transmission lines 
with a copper semi-rigid cable 
in place of the sample 
and measured the frequency dependence of the transmittance $\alpha^2(f)$ through the two cables in series at the dilution refrigerator running temperature ($\sim 100$ mK). The obtained transmission coefficient, $\alpha(f)$ is basically a monotonic decreasing function from -50.0 dB for $f < 30$ MHz down to -51.5 dB at $f=1000$ MHz with several structures which originate from the bias tee. We also measured $\alpha(f)$ at elevated temperatures and found that it was independent of temperature below 4.2 K for 1 MHz $\leq f \leq$ 1000 MHz. We examined the small difference \red{in length} between the two individual transmission lines \red{at room temperature} and estimated the error thus determining $\alpha(f)$ to be 2\%.
The input power ${\cal P}_0(f)$ to the line was adjusted to give the desired incident power ${\cal P}=\alpha^2(f){\cal P}_0(f)$ at the sample.

Since the impedance of the sample $Z$ is expected to be much larger than the line impedance $Z_0=50\,{\rm \Omega}$ for $f$ of the present measurement, 
the reflection coefficient of the RF voltage at the input edge of the sample becomes
 $\Gamma=(Z-Z_0)/(Z+Z_0)\simeq 1$.
Thus, twice the incident voltage is applied to the edge of the array, and the amplitude of the applied ac voltage becomes,
\begin{equation}
V_{\rm ac}=2\sqrt{2{\cal P}Z_0}.
\label{Vac_eq}
\end{equation}
We estimated the uncertainty in determining $V_{\rm ac}$ with this procedure to be 4\% concerning the uncertainties in $Z_0$ and in determining ${\cal P}$ \cite{estimates}.


\section{Experimental Results}\label{Results}
\begin{figure}
\begin{center}
\includegraphics[width=0.9\columnwidth,clip=true]{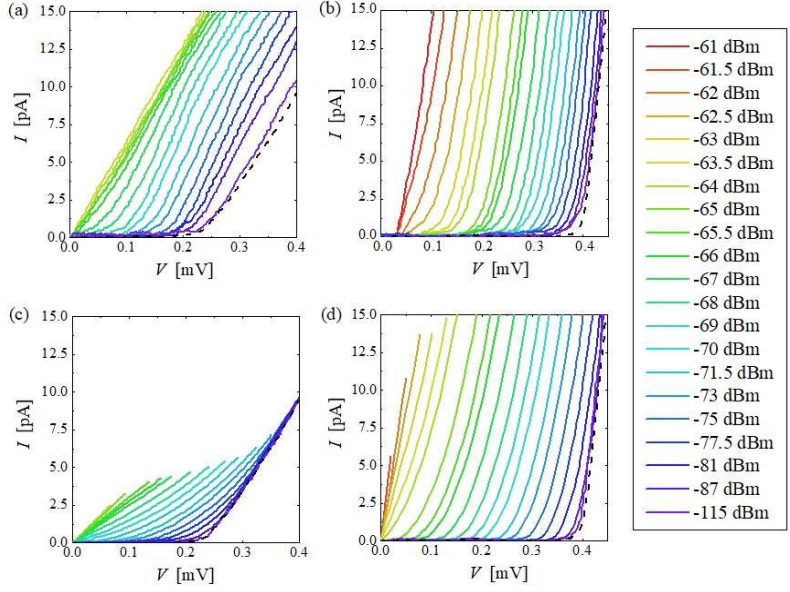}
  \caption{
  The measured $I$--$V$ characteristics of the array at 40 mK for (a) $H=0$ and (b) $H = {\rm 500} \ {\rm Oe}$. 
  Different curves correspond to different values of applied microwave power ${\cal P}$.
  The frequency of the microwave is fixed as $f = 100$ MHz. 
  The calculated $I$--$V$ curves by using Eq. (\ref{Hamilton_eq}) for (c) $H=0$ and (d) $H = {\rm 500} \ {\rm Oe}$.
  The dashed curves $I_0(V)$ in all figures correspond to measured $I$--$V$ characteristics of the array for $H$ = Oe and 500 Oe with no microwaves. 
  $V_{\rm ac}$ is calculated from ${\cal P}$ using Eq. (\ref{Vac_eq}). The horizontal dashed line indicates $I_{\rm th} = 1$ pA, which defines $V_{\rm th}$, the Coulomb blockade threshold voltage for (a)--(d) at varied values of applied microwave power ${\cal P}$.}
  \label{IVs}
\end{center}
\end{figure}
\begin{figure}
\begin{center}
\includegraphics[width=0.6\columnwidth,clip=true]{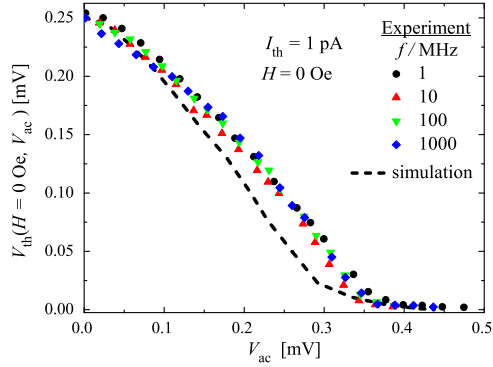}
  \caption{The Coulomb blockade voltage, $V_{\rm th}(H = 0, V_{\rm ac})$--microwave amplitude, $V_{\rm ac}$ dependence extracted from the measured $I$--$V$ characteristics for microwave frequencies 
  $f$ = 1 MHz, 10 MHz, 100 MHz and 1000 MHz, and magnetic field $H$ = 0 Oe. 
  The calculated $V_{\rm th}(H = 0, V_{\rm ac})$--$V_{\rm ac}$ dependence for magnetic field $H$ = 0 Oe is plotted as a dashed curve. 
  $V_{\rm th}(H = 0, V_{\rm ac})$ was defined at a threshold current, $I_{\rm th} = 1$ pA. }
  \label{VcbVac}
\end{center}
\end{figure}
\begin{figure}
\begin{center}
\includegraphics[width=0.7\columnwidth,clip=true]{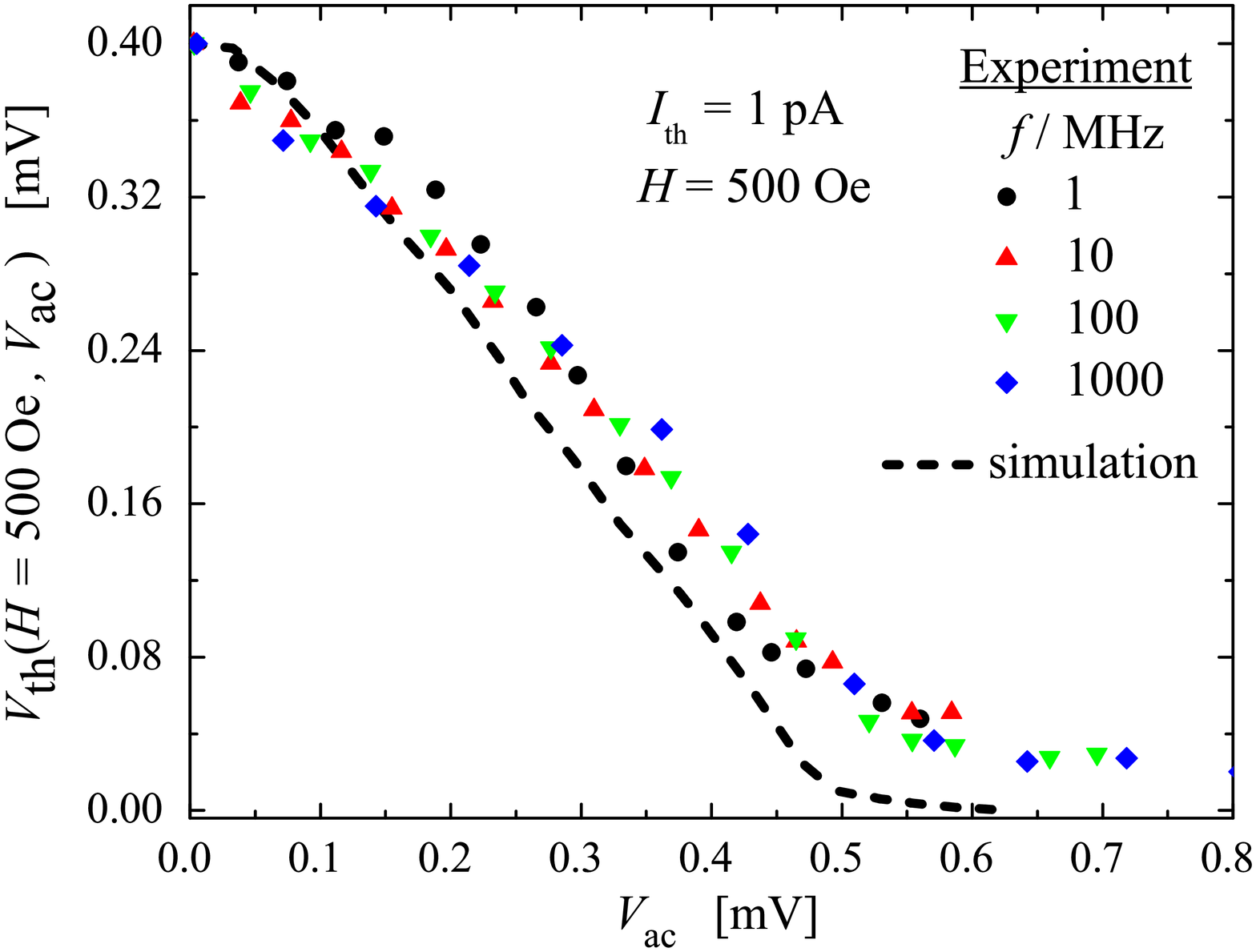}
  \caption{The Coulomb blockade voltage, $V_{\rm th}(H = 500$ ${\rm Oe}, V_{\rm ac})$--microwave amplitude, $V_{\rm ac}$ dependence extracted from the measured $I$--$V$ characteristics for microwave frequencies 
  $f$ = 1 MHz, 10 MHz, 100 MHz and 1000 MHz, and magnetic field $H$ = 500 Oe. 
  The calculated $V_{\rm th}(H = 500$ ${\rm Oe}, V_{\rm ac})$--$V_{\rm ac}$ dependence for magnetic field $H$ = 500 Oe is plotted as a dashed curve. 
  $V_{\rm th}(H = 500$ ${\rm Oe}, V_{\rm ac})$ was defined at a threshold current, $I_{\rm th} = 1$ pA. }
  \label{VcbVac500}
\end{center}
\end{figure}

We measured the $I$--$V$ characteristics of the array at 40 mK for different values of the applied RF microwave power ${\cal P}$ ranging from -115 dBm to -60 dBm. 
Figure 2(a) shows the measured $I$--$V$ characteristics at $H$ = 0\,Oe for $f$ = 100\,MHz. 
We can observe a clear Coulomb blockade for ${\cal P}=$ 0 (the dotted line).
The nonlinearity of the $I$--$V$ curves decreases as $P$ increases and exhibit nearly Ohmic behaviour at ${\cal P} \ge -63 $ dBm. 
Thus, the Coulomb blockade is gradually lifted with increase in microwave power ${\cal P}$.

We define the Coulomb blockade voltage $V_{\rm th}(H, V_{\rm ac})$ at a threshold current $I_{\rm th}$ = 1 pA in the  $I$--$V$ curves for finite microwave amplitudes $V_{\rm ac}$ calculated with Eq. (\ref{Vac_eq}). \red{For instance, consider the case with $H$ = 0 Oe and $f = 100 \ {\rm MHz}$. While the Coulomb blockade width in the absence of the RF field ($V_{\rm ac}=0$) was about $V_{\rm th}(H=0, 0) = 0.25$ mV, it was diminished to $V_{\rm th}(H=0, V_{\rm ac}) < 0.05 \ {\rm mV}$ with the application of microwave power, $V_{\rm ac} = 0.42$ mV (${\cal P}= -63.5$\,dBm)} as shown in Fig. \ref{IVs}(a) and Fig. \ref{VcbVac} (inverted triangle).
Similarly, when $V_{\rm th}(H, V_{\rm ac})$ was increased to about $V_{\rm th}(H = 500$ Oe$, 0)$ = 0.34 mV = $1.4\times V_{\rm th}(H = 0$ Oe$, 0)$ by applying an external magnetic field of $H$ = 500 Oe, 
and microwaves of the same frequency of $f$ = 100 MHz applied to the sample, Coulomb blockade was diminished with increase in microwave power as shown in Fig. \ref{IVs}(b) and Fig. \ref{VcbVac500} (inverted triangle). In this case, the Josephson coupling energy per junction was suppressed from $E_{\rm J}(H = 0)$ = 30.3 \,$\mu$eV to $E_{\rm J}(H = 500$ Oe$)=$ 24.5\,$\mu$eV by the magnetic field $H$ (as shown in Table \ref{tab:a}). 
The magnetic field value of $H = 500$ Oe was chosen at nearly maximum Coulomb blockade voltage \cite{Shimada2012}.

The aforementioned results were replicated for microwave frequencies of 1 MHz $\leq f \leq$ 1000 MHz. 
Typical results are demonstrated in Fig. \ref{VcbVac} and Fig. \ref{VcbVac500} for $f$ = 1, 10, 100, 1000 \,MHz. 

In Figs. 2-4, we also presented simulated characteristics. All the simulated values were calculated using their corresponding RF field-free curves for $f$ = 100 MHz (plotted as dotted curves) in Fig. \ref{IVs} as the function $I_{0}(V)$ in Eq. (\ref{Hamilton_eq}) as discussed in the next section.

\section{Discussion}

\red{\subsection{Considering electron heating by the RF Irradiation}}

\red{We first consider the effect of electron heating by RF irradiation of the sample before discussing the effect of the ac voltage. Dissipation in the array for the maximum applied RF (-61 dBm) is estimated to be approximately 10 fW by roughly estimating the differential resistance at the threshold range to be 10 M${\rm \Omega}$. Applying the discussion given in Refs. \cite{Korotkov1994,Nahum1994} on electron heating in the single electron transistor, the electron-phonon coupling constant 2 nW\,K$^{-5}$\,$\mu$m$^{-3}$ in Al and the volume 0.12\,$\mu$m$^3$ of our array leads to electron temperature of 130 mK. Thus, with the increase in $V_{\rm ac}$, the electron temperature rises above the substrate temperature 40 mK. However, as demonstrated in the temperature dependence of $V_{\rm th}$ in a similar Al/AlO$\rm _x$/Al-junction array in Ref. \cite{Cedergren2015}, the decrease in $V_{\rm th}$ of the array is fairly small ($\lsssim$ 10 \%) up to 130 mK. This also holds in the case of electron heating since the origin of the decrease in $V_{\rm th}$ is considered to be quasi-particle excitation in the Al electrode \cite{Cedergren2015}. Hence, the electron heating only slightly affects the obtained curves in Figs. 3 and 4 at the largest $V_{\rm ac}$ range. The number of excited quasi-particles in each electrode in the array at the electron temperature of 130\,mK are estimated to be around $0.01$ and $0.2$ for $H=$ 0\,Oe and 500\,Oe, respectively, using a Boltzmann factor, $\exp(-{\rm \Delta}/k_{\rm B}T)$ and the expression $N_{\rm eff}\sim {\cal V}\rho(0)\sqrt{2\pi k_{\rm B}T{\rm \Delta}}$ for the effective number of quasi-particle states where we have substituted density of states $\rho(0)=1.45\times 10^{47}$\,m$^{-3}$J$^{-1}$ for Al, the volume of the electrode ${\cal V}=0.013\, \mu$m$^3$ and the values of ${\rm \Delta}$ in Table I \cite{Cedergren2015}. Thus, {\it in the range of applied $V_{\rm ac}$ on our sample, the number of excited quasi-particle in the electrode can be considered negligibly small.}}

\subsection{Photon-assisted tunneling of Cooper-pairs}

The effect of an ac voltage on a single small Josephson junction was theoretically considered by Falci et al. \cite{Falci1991}, obtaining the following expression for the dc $I$--$V$ characteristics with an ac voltage superposed,
\begin{equation}
I(V)=\sum_{n=-\infty}^{\infty}J_n^2(2eV_{\rm ac}/hf)\,I_0(V-nhf/2e),
\label{TG_eq}
\end{equation}
Here, $J_n(x)$ are the Bessel functions of the first kind, and $I_0(V)$ denotes the $I$--$V$ characteristics in the absence of the RF voltage ($V_{\rm ac}=0$). 
This is the well-known Tien-Gordon formula and corresponds to the photon-assisted tunneling of Cooper pairs through the junction \cite{Tien-Gordon1963}.

In the present experiment, the frequency $f$ is sufficiently small compared to the amplitude of the ac voltage ($hf\ll eV_{\rm ac})$.
In this situation, the formula (\ref{TG_eq}) becomes \cite{Hamilton1970}
\begin{align}
I(V)=\frac{1}{\pi}\int_{-\pi/2}^{\pi/2}{\rm d}\theta \, I_0(V-V_{\rm ac} \sin \theta) .
\label{Hamilton_eq}
\end{align}
Here the nature of the photon-assisted transport is smeared out, and the current is described by the average of steady-state currents at each time.

We numerically simulated the $I$--$V$ curves for various values of $V_{\rm ac}$ using Eq. (\ref{Hamilton_eq}) and the measured $I$--$V$ curves $I_0(V)$ in the absence of the RF voltage.
The numerical results for $H = 0$ and $H = 500\ {\rm Oe}$ are shown in Fig.~\ref{IVs}~(c) and (d), respectively.
The calculated $I$--$V$ curves well reproduce the features of the measured $I$--$V$ characteristics shown in Fig.~\ref{IVs}~(a) and \ref{IVs}(b). 
This lifting of the Coulomb blockade is a dual effect to the microwave-enhanced phase diffusion observed by Liou et al. \cite{Liou2008} in a 1D array of Josephson junctions with $E_{\rm J}/E_{c}>1$.

We also extracted the Coulomb blockade voltages $V_{\rm th}$ from the calculated $I$--$V$ curves with the same criterion of $I = 1\ {\rm pA}$ as the experimental curves.
The results are plotted, together with the experiment ones, as dashed lines in in Fig. \ref{VcbVac} and Fig. \ref{VcbVac500} for $H=$ 0 Oe and 500 Oe cases respectively.
The general feature of the measured results was reproduced, albeit slightly steeper curves than the measurement results.

\subsection{Scaling analysis}

At zero temperature, the RF-free $I$--$V$ curve ($V_{\rm ac}=0$) is approximately written as
\begin{equation}
I_{0}(V)=\frac{V-V_{\rm cb}}{R^*}\Theta(V-V_{\rm cb}),
\label{eq:I0Tzero}
\end{equation}
where $V_{\rm cb}$ denotes the (ideal) Coulomb blockade voltage, $R^*$ is the differential resistance above the threshold that can depend on the magnetic field $H$, and $\Theta(x)$ is the Heaviside function. 
In experiments at finite temperatures, the RF-free $I$--$V$ curve is smeared a little from the ideal form given in Eq.~(\ref{eq:I0Tzero}).
For analysis of the experimental data, we assume a scaling form for the RF-free $I$--$V$ curve as follows:
\begin{align}
I_{0}(V) = \frac{V_{\rm cb}}{R^*} g(V/V_{\rm cb}),
\end{align}
where $R^*$ is a typical value of the differential resistance above the threshold, and $g(x = V/V_{\rm cb})$ is a universal scaling function independent of the temperature and the magnetic field \cite{scaling}.
For the ideal case satisfying Eq.~(\ref{eq:I0Tzero}), the scaling function becomes $g(x)=(x-1)\Theta(x-1)$.
In general, $g(x)$ is almost zero for $x<1$, and is a rapidly increasing function for $x>1$.
If this scaling holds, the $I$--$V$ curves fall on a unified line when the voltage and the current are normalized as $V/V_{\rm cb}$ and $IR^*/V_{\rm cb}$.
\red{The experimental data satisfies this scaling at least near the threshold region \cite{scaling2}}.
In the absence of the RF, the threshold voltage $V_{\rm th}$ is determined by the equation
\begin{align}
I_{\rm th} =  \frac{V_{\rm cb}}{R^*}  g(V_{\rm th}(V_{\rm ac}=0)/V_{\rm cb}).
\end{align}
The solution of this equation is approximately given as
\begin{align}
V_{\rm th}(V_{\rm ac}=0) \simeq V_{\rm cb},
\label{eq:VthVac0}
\end{align}
if the threshold current $I_{\rm th}$ is appropriately taken (e.g. along the red dotted line at $I_{\rm th} = 1$ pA in Fig.~\ref{IVs}).

\begin{figure}
\begin{center}
\includegraphics[width=0.6\columnwidth,clip=false]{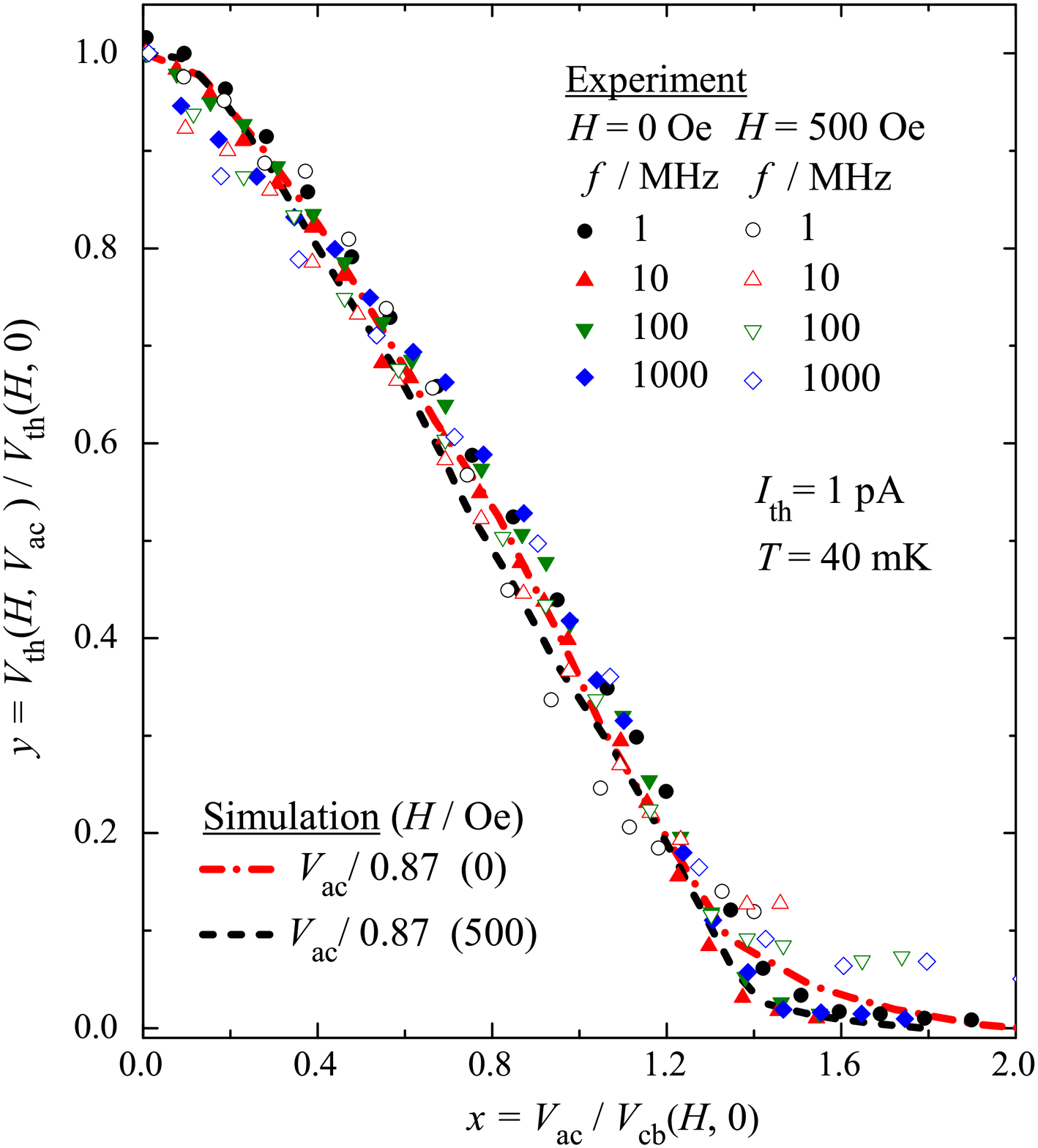}
  \caption{Normalized Coulomb blockade voltage, $V_{\rm th}(H, V_{\rm ac})/V_{\rm th}(H, 0)$--normalized RF field amplitude, $V_{\rm ac}/V_{\rm th}(H, 0)$ dependence 
  extracted from the measured $I$--$V$ characteristics for RF field frequencies $f$ = 1 MHz, 10 MHz, 100 MHz and 1000 MHz, and magnetic fields $H$ = 0 Oe and $H$ = 500 Oe. 
  $V_{\rm th}(H, 0)$ is defined at a threshold current, $I_{\rm th} = 1$ pA.  
  All curves are normalized by their corresponding $V_{\rm cb}(H, 0)$ values. 
  The normalized experimental values differ from the normalized simulated dashed curves by a factor $\Xi = $0.87.
  } 
\label{Vcb_normVac_norm}
\end{center}
\end{figure}

Using Eqs.~(\ref{Hamilton_eq}) and (\ref{eq:VthVac0}), the equation to determine the threshold voltage $V_{\rm th}$ under the RF becomes,
\begin{align}
\frac{I_{\rm th} R^*}{V_{\rm cb}} \simeq \frac{1}{\pi} \int_{-\pi/2}^{\pi/2} d\theta \, g(V_{\rm th}/V_{\rm th}(V_{\rm ac}=0)- \sin \theta V_{\rm ac}/V_{\rm th}(V_{\rm ac}=0)).
\end{align}
By solving this equation, $V_{\rm th}/V_{\rm th}(V_{\rm ac}=0)$ is obtained as a function of $I_{\rm th} R^*/V_{\rm cb}$ and $V_{\rm ac}/V_{\rm th}(V_{\rm ac}=0)$.
If the threshold current $I_{\rm th}$ is appropriately taken again, one may expect that the dependence of $I_{\rm th} R^*/V_{\rm cb}$ on the magnetic field is negligible.
Therefore, we obtain the scaling for the threshold voltage as follows:
\begin{align}
V_{\rm th}/V_{\rm th}(V_{\rm ac}=0) = h(V_{\rm ac}/V_{\rm th}(V_{\rm ac}=0) ),
\end{align}
where $h(x)$ is a scaling function independent of the magnetic field and the RF field.
Figure~\ref{Vcb_normVac_norm} is the plot of measured $y = V_{\rm th}/V_{\rm th}(V_{\rm ac}=0)$ as a function of $V_{\rm ac}/V_{\rm th}(V_{\rm ac}=0)$ for two $H$ values of 0 Oe and 500 Oe at various RF frequencies.
All the characteristics fall on the same line irrespective of $H$ or $f$ as expected, which demonstrates the validity of this scaling.
This means that $2eV_{\rm cb}$ is the dominant energy scale in the present phenomenon.

In Fig.~\ref{Vcb_normVac_norm}, we also plotted the same curve produced by numerical simulation using Eq. \ref{Hamilton_eq}, where $I_{0}$ is the $I$--$V$ characteristics when $V_{\rm ac} = 0$ and $H =$ 0 Oe and 500 Oe. The simulated curves exhibit (nearly) identical characteristics.
However, to fit them to the measured results, we need to divide $x$ by a factor $\Xi =$ 0.87. This factor is measured beyond the estimated uncertainty, 4\%, to determine $V_{\rm ac}$ in our measurement.
This means that the actual effect of the RF-microwave voltage is suppressed by this factor compared to the bare $V_{\rm ac}$. We consider this to be a renormalization effect of the RF microwave applied to the array, and we will analyze it in Sec.~\ref{sec:renormalization}.

\subsection{Magnetic field dependence}

While the magnetic field does not change the scaling $h(x)$, its effect appears in the threshold voltage $V_{\rm th}(V_{\rm ac}=0)$.
The experimental data indicates
\begin{align}
\frac{V_{\rm th}(H=500\ {\rm Oe},V_{\rm ac}=0)}
{V_{\rm th}(H=0,V_{\rm ac}=0)} \simeq 1.4
\end{align}
in our sample.
This enhancement of the Coulomb blockade is due to the suppression of the Josephson energy $E_{\rm J}$ (see Table~\ref{table:parameter}). However, it is difficult to calculate the $E_{\rm J}$-dependence of $V_{\rm th}(V_{\rm ac}=0)$, since we have to carefully consider many-body effects of electrons in the Josephson junction arrays.
A rough estimate using the concept of the depinning potential \cite{Vogt2015} gives 
\begin{align}
\frac{V_{\rm th}(H=500\ {\rm Oe},V_{\rm ac}=0)}
{V_{\rm th}(H=0,V_{\rm ac}=0)}
= \left [U(0.22)/U(0.27) \right ]^{2/3} \simeq 1.05,
\end{align}
where $U(E_{\rm J}/E_{\rm c})$ is the depining potential (see Fig. 4 of Ref. \cite{Vogt2015}).
More accurate analysis as well as detailed measurements under varied values of the magnetic field is left as the future problem.

\subsection{Renormalization factor}
\label{sec:renormalization}

\begin{figure}[tb]
\begin{center}
\includegraphics[width=0.6\columnwidth,clip=true]{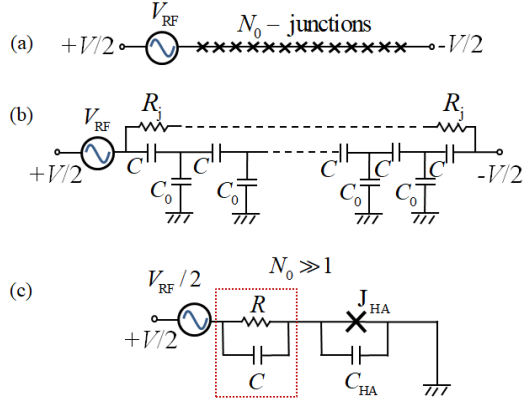}
  \caption{(a) The schematic of the symmetrically dc-biased array with identical tunnel junctions. The RF signal is applied from the left; 
  (b) The equivalent circuit model showing $R_{\rm j}$, the environmental resistance of each junction and the relevant capacitances of the array, where $C$ is the capacitance of each junction and $C_{0}$ is the stray capacitance of the electrode between the junctions. Due to the symmetric bias ($+V/2$ and $-V/2$), the dc voltage at the center electrode is assumed to be zero;
  (c) A simplified circuit of the {\it half} array (HA), with (a) and (b) the {\it whole} array, $R = N_{0}R_{j}/2$ and ${\rm J}_{\rm HA}$ act as the total environmental resistance and the {\it half} array respectively. For a large number of junctions ($N_0 \gg 1$), the combined capacitance of the {\it half} array is calculated as $C_{\rm HA} = \left ( C_0 + \sqrt{C_0^2+4CC_0} \right )/2$. The circuit elements, $R$ and $C$ within the red dotted rectangle represent the total effective impedance as seen by the {\it half} array ${\rm J}_{\rm HA}$ with an effective capacitance $C_{\rm HA}$ and effective impedance $Z(f)$, where $Z^{-1}(f) = R^{-1} + 2\pi i f C$ is the effective admittance of the {\it half} array.}
\label{equiv_circuit_infty}
\end{center}
\end{figure}

In this section, we roughly discuss the origin of the renormalization factor $\Xi$ ($=0.87$ in the present data) with the anzats that, since the Josephson coupling $E_{\rm J}$ is small, its inductive contribution to the total impedance of the array can be neglected. Moreover, since exact analysis of the threshold voltage in the Josephson junction array is difficult, we perform an approximate analysis using an effective circuit of the array following closely Ref. \cite{Cole2014}. Here, we show that the difference in the response of the dc and ac voltages can be evaluated by theoretically estimating the effective impedance of the array.

In particular, we first describe the Josephson junction array (Fig.~\ref{equiv_circuit_infty}~(a)) with an effective circuit shown in Fig.~\ref{equiv_circuit_infty}~(b). However, the expression for the effective impedance derived by employing the theory of continued fractions \cite{ContFrac2008} using this effective circuit is unwieldy; a more practical approach is to further transform the effective circuit given in Fig.~\ref{equiv_circuit_infty}~(b) into  Fig.~\ref{equiv_circuit_infty}~(c). 
Because the voltages with opposite signs ($+V/2$ and $-V/2$) are applied to the two edges of the Josephson junction array, the electric potential at the center of the array (corresponding to the voltage drop at the capacitance $C_0$) is almost zero. Therefore, the effective circuit is transformed into the equivalent circuit shown in Fig.~\ref{equiv_circuit_infty}~(c). 
Here, we have replaced a series of circuit elements in the {\it half} array with effective circuit elements. We have replaced a series of the resistance element $R_{\rm j}$ with a single resistance $R = N_0 R_{\rm j}/2$, a series of the Josephson elements with a single Josephson element ${\rm J}_{\rm HA}$, and a series of the capacitances, $C_0$ and $C$ with a single capacitance element given by 
\begin{equation}
C_{\rm HA} = \frac{1}{2} \left( C_0 + \sqrt{C_0^2+4CC_0}\right),
\end{equation}
which is an analytic expression of the {\it half} array for large $N_0$. We also define the {\it half} array effective impedance $Z(f)$ as,
\begin{align}
    Z^{-1}(f) = R^{-1} + 2\pi i f C.
\end{align}

In the absence of the RF field ($V_{\rm ac}=0$), the effective impedance $Z(f)$ becomes $R^{-1}$, and in the final steady state, the dc voltage $V$ is directly applied to the Josephson junction element ${\rm J}_{\rm HA}$ with a parallel capacitance $C_{\rm HA}$. On the other hand, under the RF field ($V_{\rm ac}\ne 0$), the effective impedance $Z(f)$ approximately becomes $1/(2\pi if C)$ when the frequency $f$ is much larger than the inverse of the time constant $RC$ \cite{frequency}.
Therefore, the effective ac voltage applied to the Josephson junction element ${\rm J}_{\rm HA}$ is given by the impedance ratio 
\begin{equation}
  \lim_{RCf \to +\infty} \frac{(2\pi if C_{\rm HA})^{-1}}{Z(f) + (2\pi if C_{\rm HA})^{-1}} 
  = \frac{C}{C + C_{\rm HA}}
\end{equation}
multiplied to the amplitude of the external ac voltage $V_{\rm ac}$. This ratio corresponds to the response of the dc and ac voltages at the center junction producing the renormalization factor
\begin{equation}
\Xi = \frac{C}{C+C_{\rm HA}} = \exp(-\Lambda^{-1}),
\end{equation}
dependent on the characteristic decay length of the electric potential in the Josephson junction array defined as \cite{Liou2008,Bakhvalov1989,Walker2015}
\begin{equation}
    \Lambda = \left[ \cosh^{-1}\left( 1+ \frac{C_0}{2C} \right) \right]^{-1} \simeq \sqrt{\frac{C}{C_0}}.
\end{equation}
For the array in the present experiment, 
we obtain $\Lambda\simeq \sqrt{C/C_0}\simeq 9$ using $C_0 \simeq 9 \ {\rm aF}$, which was measured independently on the array of the same structure using the gate effect \cite{Cedergren2017}. This leads to $\exp(-\Lambda^{-1})\simeq 0.89$, which is comparable to the value experimentally determined.

\subsection{Possible applications}

The straightforward application of the observed lifting of the Coulomb blockade is RF electromagnetic field detection. Here, we make a rough 
estimate of the small-signal sensitivity of the detection. From Fig. \ref{VcbVac}, the resolvable smallest change in $V_{\rm th}$ 
is approximately 10 $\mu$V, and the corresponding $V_{\rm ac}$ 
becomes approximately \red{40} $\mu$V, which corresponds to the input power of \red{4} pW according to eq. (\ref{Vac_eq}). Thus, the small-signal sensitivity amounts to \red{more than $10^6$} V/W, a value much higher than the sensitivity of typical diode detectors by a factor of $10^3 \sim 10^4$. This high sensitivity allows for clear-cut power detection even when the array is weakly coupled to the RF field source.

Even though it is possible to use a single small Josephson junction or a single Cooper-Pair transistor (SCPT) in place of the array as a detector, since a distinct Coulomb blockade is essential to observing the aforementioned high sensitivity, a special arrangement of the surrounding circuit is required in order to render the environmental impedance sufficiently high \cite{Watanabe2001,Watanabe2003}. This leaves the array as a potentially simpler and more practical device for highly sensitive RF electromagnetic field detection.

Furthermore, the standard way to use it is to current bias the array at pA level and monitor the decrease in voltage at 100 $\mu$V level.
Consequently, its power dissipation becomes extremely low: 0.1-1 fW, which is desirable for RF field detection in low temperatures. 
Thus, it is well suited for on-chip detection of low level RF emissions from low temperature devices.

A preliminary demonstration has been carried out successfully by detecting the microwave emission from a SCPT fabricated 
in-situ 2 $\mu$m adjacent to and deliberately decoupled from the array \cite{LT28}. Since the SCPT is decoupled from the array in contrast to Ref. \cite{Billangeon2007}, the high sensitivity was crucial to the successful detection results. Thus, the high sensitivity enables a flexible way of its use even without direct connection to the microwave source.

\section{Conclusion}
The influence of RF electromagnetic fields on the characteristics of Cooper-pair tunneling in small Josephson junctions was studied using a one-dimensional array 
of ten small Al tunnel junctions. 
In the range from 1 MHz to 1000 MHz, the Coulomb blockade was lifted gradually with increase in applied alternating voltage $V_{\rm ac}$ of the microwaves 
irrespective of its frequency $f$. 
This decrease in the Coulomb blockade threshold voltage $V_{\rm th}$ with increase in $V_{\rm ac}$ is theoretically analyzed within the Coulomb blockade framework \cite{Falci1991}. On further analysis, we find that the RF power detected by the array is less than the expected value from simulation results by a multiplicative factor (0.87), independent of the frequency and applied magnetic field as is evident in Fig. \ref{Vcb_normVac_norm}. This factor is understood to arise from the difference in the response between applied dc and ac voltages, and is determined by theoretically estimating the effective impedance of the array to yield $\Xi = \exp(-\Lambda^{-1}) \simeq 0.87$ dependent solely on $\Lambda$, the charge soliton length of the array. 

These results offer the prospect of using the observed gradual decrease in Coulomb blockade voltage with microwave power for (on-chip) detection 
of low RF electromagnetic field power.

\section*{Acknowledgements}

G. M. Kanyolo wishes to thank Dr. Titus Masese for insightful discussions and Bernard Kanyolo for help in simulations, the KDDI foundation for financial support and the members of the Shimada, Mizugaki and Kokubo laboratories at The University of Electro-Communications for their valuable suggestions. 
We appreciate the technical assistance by J. Kamekawa, H. Nishigaki, T. Suzuki and thank W. Kuo for discussion. We also thank Y. Nakamura and Y. Iwasawa for their support. This work was supported by JSPS KAKENHI Grants Number 24340067 and 18H05258. Part of this work was conducted at the Coordinated Center for UEC Research Facilities, The University of Electro-Communications, Tokyo, Japan. The stable supply of liquid helium from it is also acknowledged.


\begin{thebibliography}{99}
\bibitem{scaling}
In general, the $I$-$V$ curve may depend on the magnetic field not only via $V_{\rm cb}$ but also directly via the ratio $E_{\rm J}/E_{\rm c}$, and should be written as 
$I_0(V)=\frac{V_{\rm cb}}{R^*} g(V/V_{\rm cb},E_{\rm J}/E_{\rm c})$.
In the present experiment, however, the charging energy $E_{\rm c}$ is always much larger than the Josephson energy $E_{\rm J}$.
In this charge dominant regime, the change in the scaling function due to $E_{\rm J}/E_{\rm c}$ is expected to be neglected.
\bibitem{scaling2}
\red{In the present scaling analysis, we have assumed that under RF irradiation, the $I$-$V$ characteristics keeps a simple form with a unique characteristic voltage above which the current starts to increase rapidly.}
\bibitem{frequency}
The result in the high-frequency limit is valid for $f>f^*$, where $f^*=1/(2\pi RC)$.
If we roughly estimate $R$ from the zero-bias resistance below the threshold voltage as $R > 10^9 \ {\rm \Omega}$, we obtain $f^* < 0.2 \ {\rm MHz}$.
\end{thebibliography}
\end{document}